\documentclass[aps,preprintnumbers,prl,twocolumn,superscriptaddress]{revtex4}

\usepackage{epsfig,latexsym,cancel,amssymb,amsmath}
\usepackage{graphicx}
\usepackage{epstopdf}
\usepackage{hyperref, graphicx, slashed, xcolor, amsmath}

\allowdisplaybreaks

\definecolor{red}{rgb}{1,0,0}

\newcommand{\beq}{\begin{equation}}
\newcommand{\eeq}{\end{equation}}
\newcommand{\bea}{\begin{eqnarray}}
\newcommand{\eea}{\end{eqnarray}}
\newcommand{\nn}{\nonumber\\}

\begin{document}

\title{A unique gravitational wave signal from phase transition during inflation}

\author{Haipeng An}
\affiliation{Department of Physics, Tsinghua University, Beijing 100084, China}
\affiliation{Center for High Energy Physics, Tsinghua University, Beijing 100084, China}
\author{Kun-Feng Lyu}
\affiliation{Department of Physics, the Hong Kong University of Science and Technology, Clear Water Bay,
Kowloon, Hong Kong S.A.R., P.R.C.}
\affiliation{Kavli Institute for Theoretical Physics, University of California, Santa Barbara, CA 93106, USA}
\author{Lian-Tao Wang}
\affiliation{Department of Physics and Enrico Fermi Institute, University of Chicago, Chicago, IL 60637, USA}
\affiliation{Kavli Institute for Cosmological Physics, University of Chicago, Chicago, IL 60637, USA}
\author{Siyi Zhou}
\affiliation{The Oskar Klein Centre for Cosmoparticle Physics \& Department of Physics, Stockholm University, AlbaNova, 106 91 Stockholm, Sweden}

\begin{abstract}

We study the properties of the gravitational wave (GW) signals produced by first order phase transitions during the inflation era. We show that the power spectrum of the GW oscillates with its wave number. This signal can be observed directly by future terrestrial and spatial gravitational wave detectors and through the B-mode spectrum in CMB. This oscillatory feature of GW is generic for any approximately instantaneous sources occurring during inflation, and is distinct from the GW from phase transitions after the inflation. The details of the GW spectrum contain information about the scale of the phase transition and the later evolution of the universe. 

\end{abstract}

\maketitle

\section{Introduction}

Gravitational waves (GWs), once produced, propagate freely through the universe and can bring us the information of their origin and the history of the universe. They can be detected in many proposed, either terrestrial or space based, detectors~\cite{Seoane:2013qna,Audley:2017drz,Kawamura:2011zz,Luo:2015ght,Guo:2018npi,Crowder:2005nr,Harry:2006fi,Corbin:2005ny,Kramer:2013kea,Hobbs:2009yy,Janssen:2014dka,TheLIGOScientific:2014jea,Abramovici:1992ah,TheVirgo:2014hva,Punturo:2010zz,Reitze:2019iox}. Primordial GWs can also leave hints on the cosmological microwave background (CMB) and can be detected in the B-mode power spectrum~\cite{Hui:2018cvg,Li:2017drr,Abazajian:2019eic}. Possible sources of the primordial GWs are inflation~\cite{Grishchuk:1974ny,Starobinsky:1979ty,Rubakov:1982df,Fabbri:1983us,Abbott:1984fp}, first order phase transitions~\cite{Witten:1984rs,Kamionkowski:1993fg}, and cosmic strings~\cite{Vachaspati:1984gt,Brandenberger:1986xn,Hindmarsh:1990xi,Damour:2001bk,Siemens:2001dx,Hindmarsh:1994re}. 

It is highly plausible that there was an inflationary era in early universe~\cite{Guth:1980zm,Linde:1981mu,Albrecht:1982wi} (See Ref.~\cite{Baumann:2009ds}).
The simplest inflation model is driven by a slow rolling inflaton. To produce enough inflation, 
the typical excursion of the inflaton field must be large. As such, it may induce significant changes in the dynamics of the spectator fields. 
This may happen through a direct coupling between the inflaton field to other spectator fields (see, e.g., Ref.~\cite{Chen:2009zp}). The change in temperature during inflation is another possibility (see Refs.~\cite{Berera:1995wh,Berera:1995ie} as examples). Such changes can trigger dramatic events during inflation, such as a first order phase transition~\cite{Jiang:2015qor,Wang:2018caj}. 
The inflation era may also start from a first order phase transition~\cite{Sugimura:2011tk}.

In this letter, we show that the GWs produced by bubble collisions in first order phase transition during inflation can provide a unique {\it oscillatory} signal in its power spectrum, which contains information of both inflation and the phase transition. It should be clear from the discussion below that the signal is generic for approximately instantaneous GW sources. 


{\noindent \bf GWs from instantaneous sources.} The equation of motion for the transverse and traceless GW perturbation $h_{ij}$ is 
\bea\label{eq:wave}
h''_{ij} + \frac{2a'}{a} h'_{ij}- \nabla^2 h_{ij} = 16 \pi G_N a^2 \sigma_{ij} \ ,
\eea
where $'$ indicates derivatives with respect to the conformal time $\tau$, $G_N$ is the Newton's gravity constant, and  $\sigma_{ij}$ is the transverse, traceless part of the energy momentum tensor. In this work, we assume the Hubble parameter, $H_{\rm int}$, during inflation is a constant. Then, we have $a(\tau) = - 1/H_{\rm inf} \tau$. 
There are several important time scales in the problem: $\tau_\star$ is the time of bubble collision and generation of the GW. The inflation ends at $\tau_{\rm end}$. We denote the conformal time duration and the co-moving spatial spread of the bubble collision event to be $\Delta_{\tau,x}$. $\Delta_{\tau,x} \ll |\tau_\star|$ by assumption for instantaneous and local sources that happened during inflation. 
 The modes of interest to us are all outside the horizon at the time when inflation ends, $k|\tau_{\rm end}| \ll 1 $, where $k$ is the co-moving momentum. 

%

\section{Spectral shape of the GW signal}
We focus on three regimes with qualitatively different features. 

\bigskip

{\noindent $\bullet \  |\tau_\star|^{-1} < k < \Delta_{\tau,x}^{-1}$.} 
In this regime, we can ignore the spatial inhomogeneity caused by the bubbles and treat the bubble collisions as instantaneous sources. Therefore the bubble collisions can be approximated as delta function sources, 
\bea\label{eq:source}
\tilde\sigma_{ij} \approx \tilde T_{ij}^{(0)}a^{-3}(\tau_\star) \delta(\tau - \tau_\star) \ ,
\eea
where in this regime $\tilde T_{ij}^{(0)}$ is independent of $\bf k$~\cite{Cai:2019cdl}.

\begin{figure}
\centering
\includegraphics[height=1in]{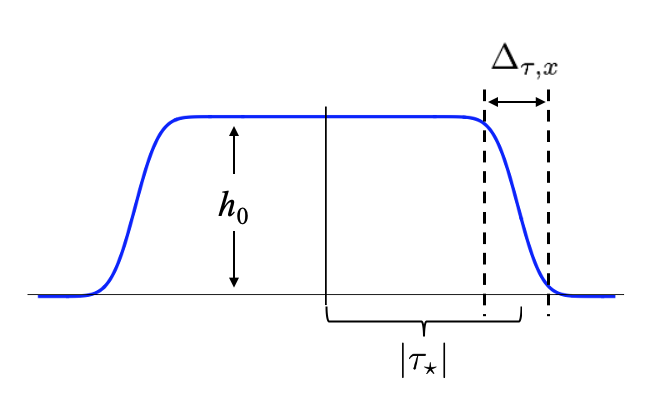}
\includegraphics[height=0.9in]{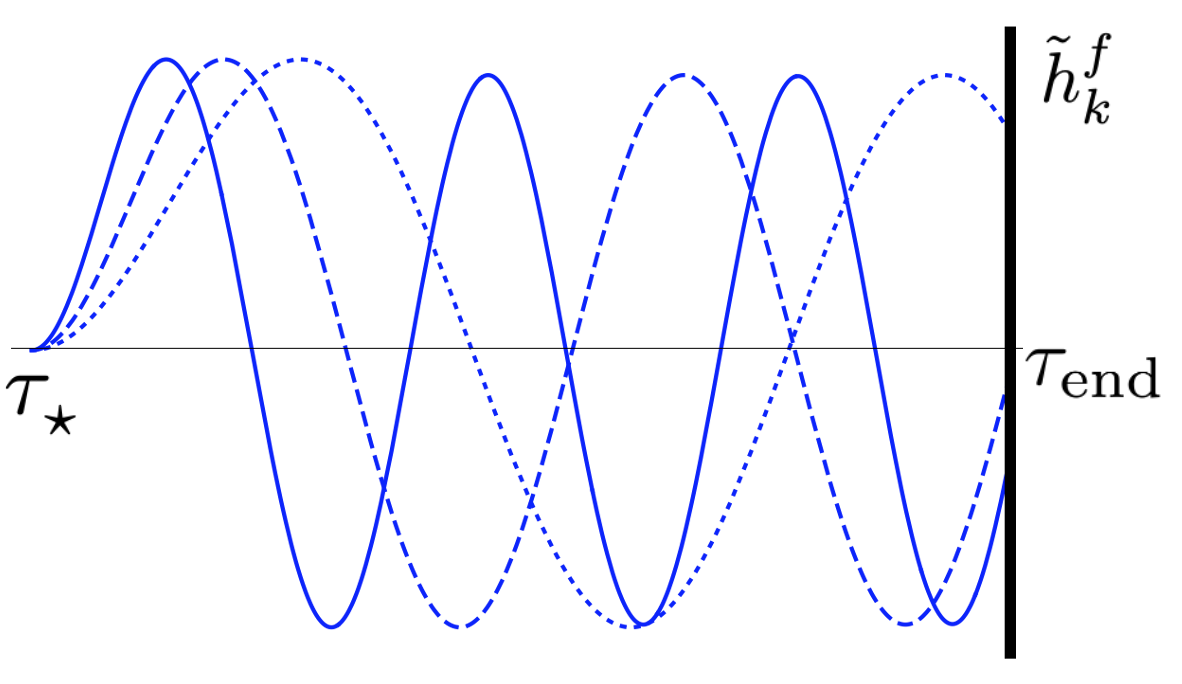}
\caption{Left: configuration of bound state of GW in the inflation era. Right: Fourier modes of GW during inflation from $\tau_\star$ to $\tau_{\rm end}$.}\label{fig:balls}
\end{figure}

During inflation and after the bubble collision, $|\tau_\star| 
\gg |\tau |> |\tau_{\rm end}|$, we have (after Fourier transformation, and suppressed $i,j$ indices )
\bea\label{eq:h_inf}
\tilde{h}_{\bf k} (\tau) & \approx & -\frac{16\pi G_N H_{\rm inf} \tilde T^{(0)} \tau }{k} \left[ \left( \frac{1}{k \tau } - \frac{1}{k \tau_\star} \right) \cos k(\tau-\tau_\star) \right. \nonumber \\  &&+  \left.\left(1+ \frac{1}{k^2(\tau \tau_\star)} \right) \sin k(\tau - \tau_\star )  \right] \ .
\eea
In position space, when $k^2 \tau\tau_\star \ll 1$, $h(\tau,{\bf x})$ is approximately $4 G_N H_{\rm inf} \tilde T^{(0)}|\tau_\star|^{-1}  \Theta(\tau -\tau_\star - |{\bf x}-{\bf x_\star}|)$, which is a uniform over density ball with radius $|\tau_\star|$, as shown in the left panel of Fig.~\ref{fig:balls}. At the end of inflation, the universe is filled with such GW balls. 
In this regime, we can ignore the terms suppressed by $(k \tau_\star)^{-1}$ in Eq.~(\ref{eq:h_inf}). After the production of the GW at $\tau \approx \tau_\star$, it will continue to oscillate until it exits the horizon at $k |\tau| \approx 1$ when its phase starts to freeze. The GW then evolves to the end of the inflation, and with $k\tau_{\rm end} \rightarrow 0$, its value is frozen to
\bea\label{eq:hf}
\tilde h_{\bf k}^{f} = - \frac{16\pi G_N H_{\rm inf} \tilde T^{(0)}}{k^2} \left(\cos k\tau_\star - \frac{\sin k\tau_\star}{k \tau_\star}\right) \ .
\eea
Since $k |\tau_\star| > 1$, we can neglect the term proportional to $\sin k\tau_\star/k\tau_\star$. Hence, $\tilde h^f_{\bf k} \sim \cos k \tau_\star$, as shown in the right panel of Fig.~\ref{fig:balls}.

The GW starts to oscillate again after re-entering the horizon, with $\tilde{h}_{\bf k}^f $ as the initial condition. For example, if we assume the universe evolves into the radiation domination (RD) immediately after inflation, 
\bea
\tilde h_{\bf k}(\tau) = \tilde h^f_{\bf k} \times \frac{\sin k\tau}{k\tau} \ .
\eea
Hence, $\tilde{h}_{\bf k}(\tau) \propto \cos (k \tau_\star)$. The energy density of the GW has the form
\bea
\rho_{\rm GW} \sim \frac{1}{a^2(\tau)} \int \frac{d^3k}{(2\pi)^3} |\tilde h'_{\bf k}(\tau)|^2 \ .
\eea
As a result, deeply inside the horizon ($k\tau \gg 1$), we have 
\bea
\label{eq:rho_osc}
\frac{d\rho_{\rm GW}}{d\log k} \sim \frac{1}{k} \left(\cos k \tau_\star - \frac{\sin k\tau_\star}{k \tau_\star} \right)^2 \approx \frac{1}{k} \cos^2 k \tau_\star \ .
\eea
 
\begin{figure}
\centering
\includegraphics[height=1.5in]{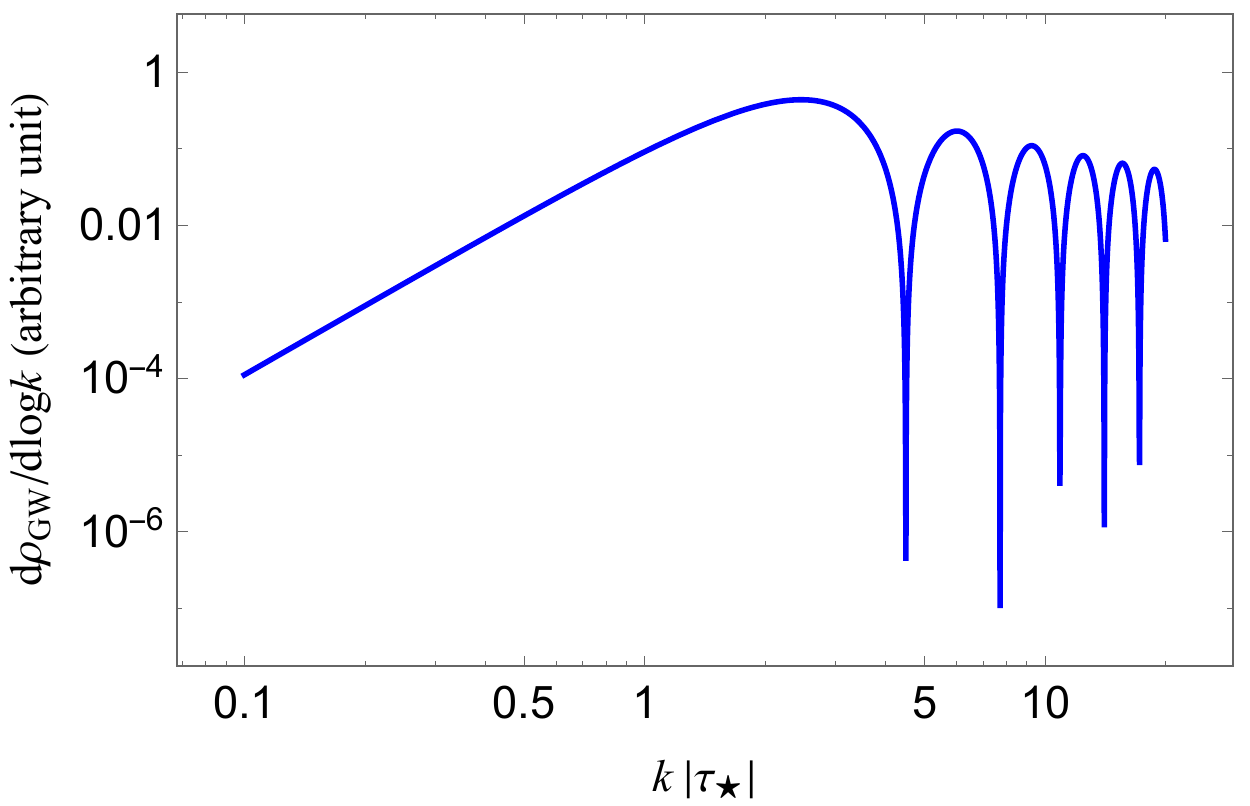}
\caption{Illustration of the shape of the GW spectrum produced by instantaneous source during inflation. }\label{fig:feature}
\end{figure}


We see that the GW has a distinct {\it oscillatory} feature in the frequency space, with a period of $\pi / \tau_\star$. This feature stems from the instantaneous nature of the GW production, which sets up a GW spectrum proportional to $\cos k|\tau_\star|$ at the end of the inflation. The GW energy density also has an overall factor of $k^{-1}$ \cite{Wang:2018caj}, since the modes with longer wavelength redshift less before exiting the horizon. An illustration of the GW spectrum is shown in Fig.~\ref{fig:feature}. 

\bigskip

{\noindent $\bullet \ k <\tau_{\star}^{-1}$. }
In this regime, we can ignore the details of the GW source, and treat it as a delta function in space-time. Hence,  Eq.~(\ref{eq:h_inf}) still applies. In the limit $k|\tau_{\rm end}| \ll k|\tau_\star| \ll 1$, from Eq.~(\ref{eq:hf}),  $h_f^k$ is independent of $k$ at leading order. From Eq.~(\ref{eq:rho_osc}) we have $d \rho_{\rm GW} /d \log k \propto k^3$, also shown in Fig.~\ref{fig:feature}, which is similar to the case of producing GW from an instantaneous source in RD~\cite{Caprini:2009fx,Cai:2019cdl}. 

\bigskip

{\noindent $\bullet \ k \gtrsim \Delta_{\tau,x}^{-1}$. } In this regime, we have $k |\tau_{\star}| \gg 1$. The details of the bubble collision become essential, and we will need  numerical simulations to obtain the shape of the signal. At such small scales, the curvature of the space-time is not important when the GW is produced. However, the inflation effect distorts the GW spectrum. As a result, the energy density behaves as 
\bea\label{eq:small}
\frac{d\rho_{\rm GW}}{d\log k} \sim k^{-4} \frac{d\rho_{\rm GW}^{\rm flat}}{d\log k_p}  \ ,
\eea
where $k_p = k/a$, is the physical momentum. ${d\rho_{\rm GW}^{\rm flat}}/{d\log k_p}$ is the GW spectrum produced from the same source in the Minkowski space-time. The distortion factor $k^{-4}$ stems from the $k^2$ factor in the denominator of Eq.~(\ref{eq:hf}). ${d\rho_{\rm GW}^{\rm flat}}/{d\log k_p}$ usually decreases as $k_p^{-r}$, with $r = 1$ for bubble collisions~\cite{Huber:2008hg}. 
Therefore, for GW produced by approximately instantaneous sources during inflation, the UV part of the spectrum decreases as $k^{-5}$. 

Due to the finite duration (of  $ {\mathcal O}\Delta_{\tau,x}$) of the sources, the oscillatory pattern in the UV part, $k \gtrsim \Delta_{\tau,x}^{-1}$ would be smeared out. %
This finite size effect should also blunt the oscillation pattern in the regime $|\tau_\star|^{-1} < k < \Delta^{-1}_{\tau,x}$. Detailed simulation can determine precisely how the spectrum is smeared. For an observer in today's universe, the GWs originate from different directions correspond to uncorrelated sources during inflation. Thus,  we can simply add up their strengths. Therefore, we can use a window function to mimic this effect by replacing the factor $(\sin k\tau_\star/k\tau_\star - \cos k\tau_\star)^2$ in Eq.~(\ref{eq:rho_osc}) with $(2\Delta)^{-1}\int_{\tau_\star-\Delta}^{\tau_\star+\Delta} d\tau_\star'(\sin k\tau_\star'/k\tau_\star' - \cos k\tau_\star')^2$, where $\Delta\sim\Delta_{\tau,x}$ embodies the duration of the source.




Combining the above analysis, the general form of the GW spectrum when it is back into the horizon in RD can be written as  
\bea\label{eq:general}
\frac{d\rho_{\rm GW}}{d\log k} = \frac{a^4(\tau_{\rm end})}{a^4(\tau)} {\cal S}(k_p) \frac{d\rho_{\rm GW}^{\rm flat}}{d\log k_p}  \ ,
\eea
where ${\cal S}(k_p)$ is 
\begin{widetext}
\bea\label{eq:S}
{\cal S}(k_p) = \frac{H_{\rm inf}^4}{k_p^4}\left\{\frac{1}{2}\! +\! \frac{\cos(2k_p/H_{\rm inf}) \sin(2k_p \Delta_p)}{4k_p \Delta_p}\!+\! \frac{1}{4k_p \Delta_p} \left( \frac{1 \!-\! \cos(2k_p/H_{\rm inf} \!-\! 2k_p \Delta_p)}{k_p/H_{\rm inf} \!-\! k_p \Delta_p} \!-\! \frac{1 \!-\! \cos(2k_p/H_{\rm inf} \!+\! 2k_p \Delta_p)}{k_p/H_{\rm inf} \!+\! k_p \Delta_p}\right)\right\}.
\eea
\end{widetext}
$\Delta_p = a^{-1}(\tau_\star) \Delta $ is the physical duration of the source. 

\bigskip

\section{Detectability of the GW signal}

To get the strength and the frequency of the GW today, we need to study a specific model.  We assume that the backreaction from the spectator sector that underwent the phase transition to the evolution of the inflaton field is negligible. To have a detectable signal, the latent heat density released during the phase transition should be  larger than $H_{\rm inf}^4$. Hence, the plasma, with energy density that can be estimated as $T_{\rm GH}^4 \sim H_{\rm inf}^4/(2\pi)^4$, is  negligible. As a result, the production of the GW is dominated by bubble collisions.  A comprehensive description of bubble collision and GW production can be found in Ref.~\cite{Huber:2008hg}. During bubble collision the parameter $\beta \equiv - d S_4/dt$ determines the size of the bubble and the wavelength of the GW, where $S_4$ is the action of the bounce at the end of the phase transition. Here we use $S_4$ since
the phase transition rate is dominated by the quantum tunneling. For the phase transition to complete during inflation, we assume $\beta \gg H_{\rm inf}$. When $\beta \approx H_{\rm inf}$, numerical simulation is needed which is beyond the scope of this work. The discussions of models in which first order phase transition can occur during inflation and the possible range of $\beta/H$ are presented in the appendix. 

In the case of an instantaneous reheating and followed by RD,   all the energy of the inflaton field converts into the radiation energy. Hence, today's relative abundance of GW can be written as
\bea\label{eq:Omega}
\Omega_{\rm GW}(k_{\rm today}) = \Omega_R \times {\cal S}(k) \times \frac{\Delta\rho_{\rm vac}}{\rho_{\rm inf}} \frac{d\rho^{\rm flat}_{\rm GW}}{\Delta\rho_{\rm vac} d\log k_p} \ ,
\eea
where $\Omega_R$ is today's abundance of radiation, $\Delta\rho_{\rm vac}$ is the latent energy density of the phase transition sector. The last factor is the flat space-time spectrum of GW
\bea\label{eq:rho0}
\frac{d\rho^{\rm flat}_{\rm GW}}{\Delta\rho_{\rm vac} d\log k_p} = \kappa^2 \left(\frac{H_{\rm inf}}{\beta}\right)^2 \Delta(k_p/\beta) \ ,
\eea
where $\kappa = 1$ if  the energy density of the plasma is negligible, and the simulation result shows
\bea
\Delta(k_p/\beta) = \tilde\Delta \times \frac{3.8\tilde k_p k_p^{2.8}}{\tilde k_p^{3.8} + 2.8k_p^{3.8}} \ ,
\eea
where $\tilde k_p = 1.44\beta$ and $\tilde \Delta = 0.077$. In the calculation of the GW spectrum, we use $\beta^{-1}$ to estimate $\Delta_p$ in ${\cal S}$. The signal strength is suppressed by the factor $(H_{\rm inf}/k_p)^4$ due to the dilution during inflation as shown in Eq.~(\ref{eq:S}). A qualitative understanding of this factor can be found in the appendix. 

\begin{figure*}
\centering
\includegraphics[height=2.5in]{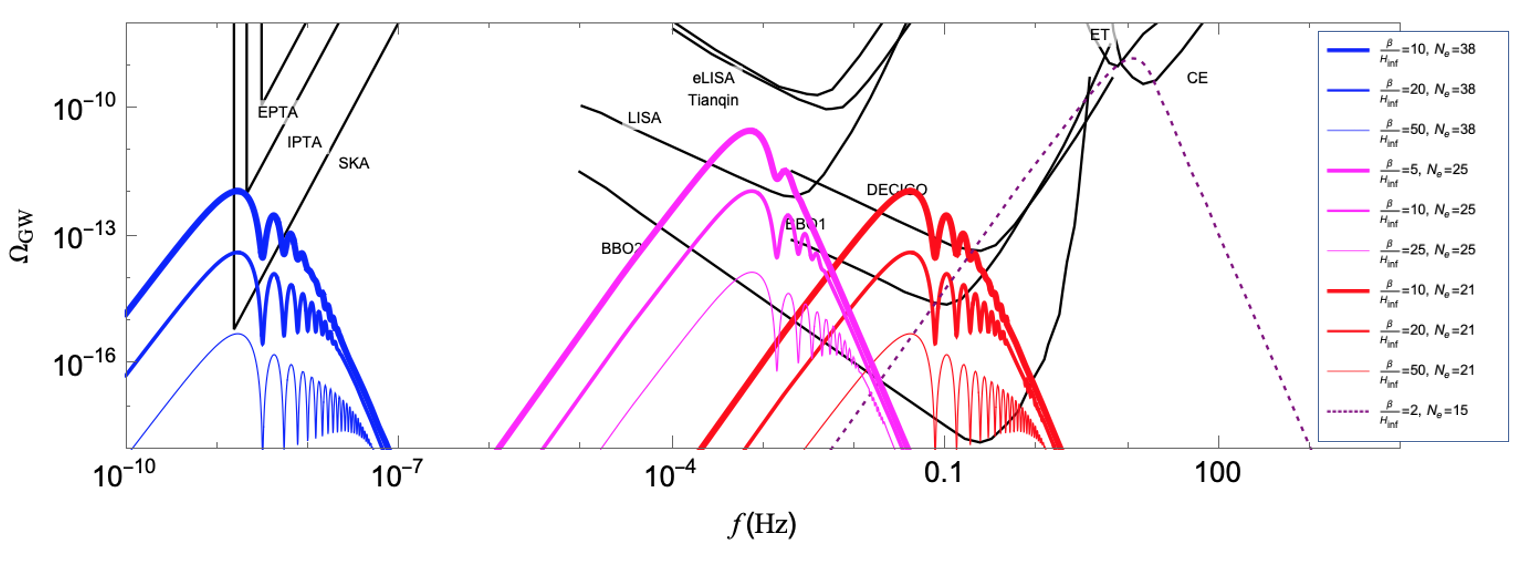}
\caption{$\Omega_{\rm GW}$ as a function of $f$. The blue, magenta, red and purple curves are for $N_e = 38$, 25, 21 and 15, respectively. The value of $\beta/H_{\rm inf}$ for each curve are shown in the legend. For all the curves $H_{\rm inf} = 10^{12}$ GeV and $\Delta\rho_{\rm vac}/\rho_{\rm inf} = 0.1$. The curve for the sensitivity of BBO phase 2 (BBO2) is from \cite{Harry:BBO2}. Curves for sensitivities of other detectors are from \cite{Moore:2014lga}.  }\label{fig:master}
\end{figure*}

Finally, the observed GW frequency is
\bea
\frac{f_{\rm today}}{f_\star} = \frac{a(\tau_\star)}{a_1} \left(\frac{g_{*S}^{(0)}}{g_{*S}^{(R)}}\right)^{1/3} \frac{T_{\rm CMB}}{\left[\left(\frac{30}{g_*^{(R)} \pi^2}\right) \left( \frac{3 H_{\rm inf}^2}{8\pi G_N} \right)\right]^{1/4}} \ ,
\eea
where the superscript $(R)$ indicates that the values of parameters at reheating temperature. 
Due to the distortion induced by inflation, the position of the highest peak of the spectrum corresponds to $k_p \approx H_{\rm inf}$. As a result, 
assuming $g_{*S}^{(R)} = g_*^{(R)} \approx 100$, the frequency of highest peak today is 
\bea\label{eq:f}
\tilde f_{\rm today} = 1.1\times10^{11} {~\rm Hz} \left(\frac{H_{\rm inf}}{m_{\rm pl}}\right)^{1/2} \left(\frac{a_\star}{a_1} \right)  .
\eea
Take high scale inflation as an instance, $(H_{\rm inf}/m_{\rm pl})^{1/2} \sim 10^{-3}-10^{-4}$. 
Detectors based on the pulsar timing technology, such as EPTA~\cite{Kramer:2013kea}, IPTA~\cite{Hobbs:2009yy}, and SKA~\cite{Janssen:2014dka} are sensitive to GWs with frequencies around $10^{-8}$ Hz. 
From Eq.~(\ref{eq:f}), they can probe the GWs produced at the era of about 40 e-folds before the end of inflation as shown by the blue curves in Fig.~\ref{fig:master}. The space-based detectors, such as LISA~\cite{Audley:2017drz}, eLISA~\cite{Seoane:2013qna}, DECIGO~\cite{Kawamura:2011zz}, 
BBO~\cite{Harry:2006fi,Corbin:2005ny}, ALIA~\cite{Crowder:2005nr}, TianQin~\cite{Luo:2015ght} and Taiji~\cite{Guo:2018npi}, is sensitive to frequencies around $10^{-2}$ Hz, 
corresponding to about 20 e-folds before the end of inflation as shown by the magenta and red curves in Fig.~\ref{fig:master}. The proposed ground-based detectors (e.g. the Einstein Telescope~\cite{Punturo:2010zz} and the Cosmic Explorer~\cite{Reitze:2019iox}) are sensitive to GWs with frequencies around $10-10^4$ Hz, which correspond to about 15 e-folds from the end of inflation. They can detect the signal of the phase transition if $\beta \approx 2 H_{\rm inf}$. However, as shown by the purple dotted curve in Fig.~\ref{fig:master}, the oscillatory feature is expected to be smeared out since $\Delta_p \sim H_{\rm inf}^{-1}$. 

\begin{figure}
\centering
\includegraphics[height=1.9in]{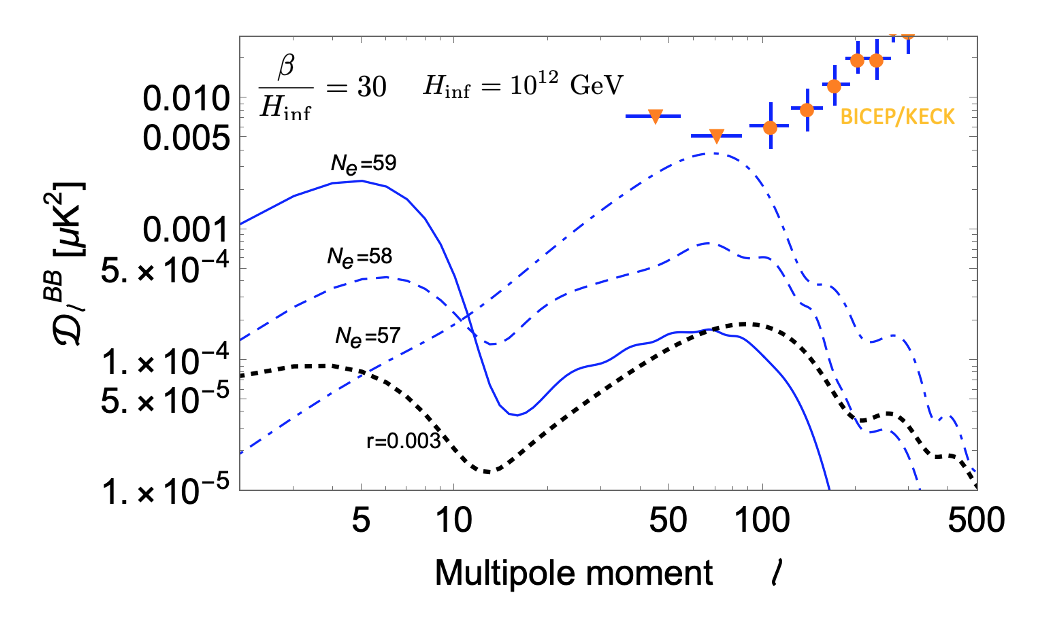}
\caption{B-mode power spectra from the bubble collision happened at $N_e = 59, 58, 57$ during inflation with $\beta / H_{\rm inf}=30$, $H_{\rm inf}=10^{12}$ GeV. $\Delta\rho_{\rm vac}/\rho_{\rm inf}=0.01$. The spectrum generated from quantum fluctuations for tensor-scalar ratio $r=0.003$ is also shown. The orange dots and downward triangles are data from the BICEP/Keck array~\cite{Array:2015xqh}. }\label{fig:Bmode}
\end{figure}

\begin{figure}
\centering
\includegraphics[height=2in]{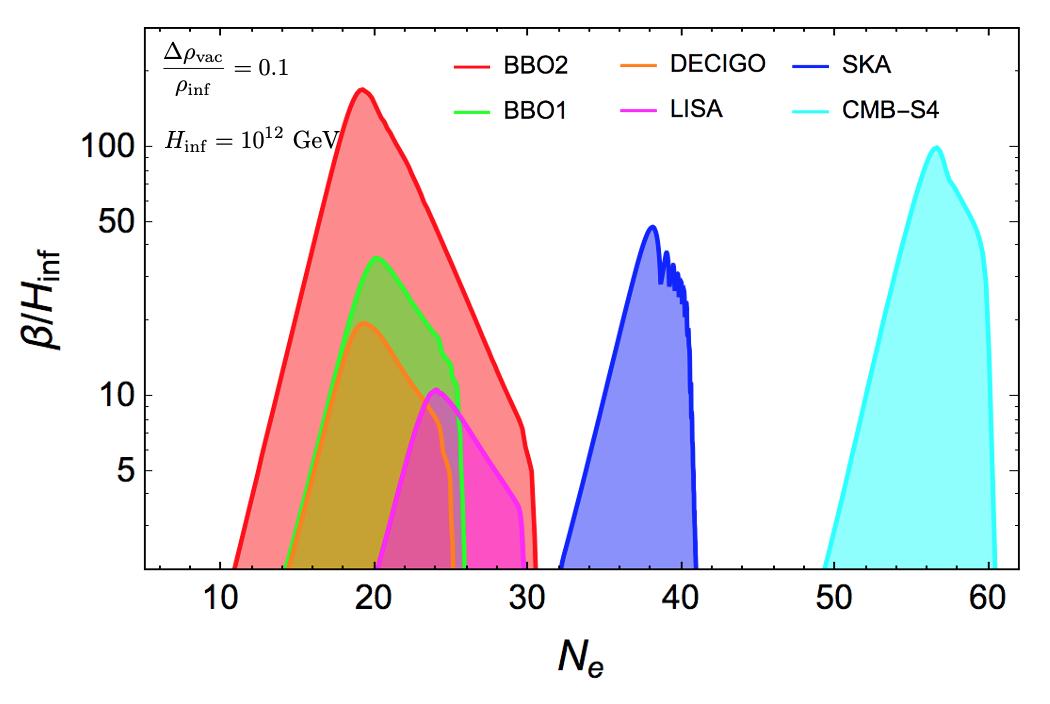}
\caption{Reaches of the proposed DECIGO~\cite{Kawamura:2011zz}, BBO~\cite{Harry:2006fi}, SKA~\cite{Janssen:2014dka} and CMB-S4~\cite{Abazajian:2019eic} projects. For CMB-S4, we require the sum of values of $D_\ell^{BB}$ of our spectrum to be smaller than the 0.003 quantum fluctuation spectrum for $\ell$ from 50 to 100. For other projects, the reach is set by requiring our signal to be below their sensitivity curves as shown in Fig.~\ref{fig:master}. }\label{fig:reach}
\end{figure}

\bigskip
\section{Signals on CMB}
If the phase transition happened about 60 e-folds before the end of inflation, it would leave an imprint on the CMB B-mode power spectrum~\cite{Jiang:2015qor}. 
Since the strength of the GWs depends on $H_{\rm inf}$ only through the ratio $H_{\rm inf}/\beta$, as shown in Eqs.~(\ref{eq:Omega}) and (\ref{eq:rho0}), it is possible to  see sizable B-mode spectrum from CMB even in low scale inflation models.

We simulate the B-mode power spectra induced by first order phase transitions using the \textsc{class} package~\cite{Blas:2011rf}. The result is shown in Fig.~\ref{fig:Bmode}, where $\beta/H_{\rm inf}$ and $\Delta\rho_{\rm vac}/\rho_{\rm inf}$ are fixed to be $30$ and $0.01$. 
The frequency of GW depends on $H_{\rm inf}$, which we have chosen to be $H_{\rm inf} = 10^{12}$ GeV. The solid, dashed and dot-dashed curves are the spectrum for $N_e$ = 59, 58 and 57, respectively.  There are small wiggles induced by the oscillatory pattern in the GW power spectrum. 
Since the spherical harmonics are not orthogonal to the Fourier modes, the oscillatory pattern is smeared. The amplitude of the oscillation is only about 10\% of the total. 
As a comparison, the black dotted curve in Fig.~\ref{fig:Bmode} shows the B-mode power spectrum produced by quantum fluctuations during inflation with the tensor-scalar ratio $r = 0.003$, which can be reached by the CMB-S4 at $5\sigma$ level~\cite{Abazajian:2019eic}. 
Of course, the inflationary history in this era will also be probed and potentially constrained by other CMB and large scale structure observables. Search for the GW signgal discussed here will provide complementary information.  We will leave a more detailed discussion to a separate work. 

\bigskip

\section{Summary and outlook}
The GW spectrum produced from instantaneous sources during inflation has an oscillatory feature, as shown in Figs.~\ref{fig:master} and \ref{fig:Bmode}, and can be detected by future GW detectors. This feature allows us to distinguish it from GW generated by sources after the inflation. From the frequency of the oscillation in the spectrum, we can learn the energy scale of the phase transition in the unit of the Hubble expansion rate during inflation. The information of the time the phase transition happened are encoded in the frequency of the GW. 
Fig.~\ref{fig:reach} shows the future reaches of LISA, DECIGO, BBO, SKA and CMB-S4 projects for $\Delta\rho_{\rm vac}/\rho_{\rm inf} = 0.1$ and $H_{\rm inf} = 10^{12}$ GeV.  

For the inflationary history outside the ten e-folds around the CMB era, there is no direct measurement of the power spectrum. Hence, the evolution there could be very different from the simple form assumed in this paper. 
At the same time, if there is a first order phase transition happened at around $N_e \geq 10$, the details of the oscillatory spectrum can help us map out this part of ``missing history". A detailed study of this subject will be presented in a separate work. 

If the phase transition happened in the regime that can be detected in CMB, the mass of the fields in the spectator sector must be larger than $H_{\rm inf}$ so that their perturbations induced by the phase transition will decay quickly after evolving out of the horizon. On the other hand, if the phase transition happens in the missing history and light degrees of freedom exist in the spectator sector, the perturbations may induce primordial black holes or dark blobs, leading to additional signals in the future.

\bigskip

\section{Appendix}

\subsection*{Models for first order phase transitions during inflation}

\begin{figure*}
\centering
\includegraphics[height=1.4in]{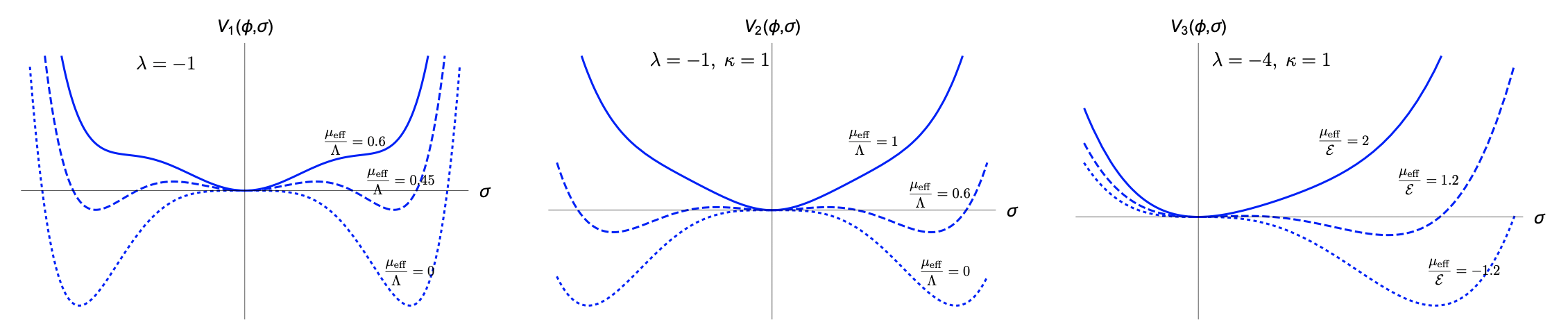}
\caption{The shapes of $V_1(\phi,\sigma)$ as a function of $\sigma$ for different chi}\label{fig:A1}
\end{figure*}

In this section, we provide some simple examples that first order phase transition can happen during inflation. As we discussed in the main text, the general scheme is that the first order phase transition happens in a spectator sector, which for simplicity, we take to be a scalar field $\sigma$. We consider the following examples of the spectator potential together with a coupling to the inflaton field $\phi$
\bea
\label{eq:model}
V_1(\phi,\sigma) &=& -\frac{1}{2} (\mu^2 - c^2 \phi^2 ) \sigma^2 + \frac{\lambda}{4}\sigma^4 + \frac{1}{8 \Lambda^2 } \sigma^6 \nn
V_2(\phi,\sigma) &=& -\frac{1}{2} (\mu^2 - c^2 \phi^2 ) \sigma^2 + \frac{\lambda}{4}\sigma^4 + \frac{\kappa}{4} \sigma^4\log\frac{\sigma^2}{\Lambda^2} \nn
V_3(\phi,\sigma) &=& -\frac{1}{2} (\mu^2 - c^2 \phi^2) \sigma^2 + \frac{\lambda}{3}{\cal E}\sigma^3 + \frac{\kappa}{4} \sigma^4 \ .
\eea
We assume during inflation, the field value of $\phi$ becomes smaller. For $\mu^2 > 0$, $\lambda < 0$ and $c^2 >0$, first order phase transition can happen. In the models in Eq.~(\ref{eq:model}), we can define an effective mass square $\mu_{\rm eff}^2 \equiv - (\mu^2 - c^2 \phi^2)$. With the rolling the inflaton field, the effective mass evolves from positive to negative. With $\lambda<0$, there would be a barrier in the potential. Hence, the phase transition would be first order. Evidently, first order phase transition can happen in all three models, as shown in Fig.~{\ref{fig:A1}}. A detailed analysis of the potentials is in parallel to the electroweak phase transition with the temperature $T$ replaced by $c\phi$, and can be found in Ref.~\cite{Chung:2012vg}.
There can also be models in which the evolution of the inflaton field changes the values of the couplings in the spectator sector. For example, 
\bea
{\cal L}_\sigma = - \left(1 - \frac{\phi^2}{\Lambda^2}\right)\frac{1}{4g^2} G^a_{\mu\nu} G^{a\mu\nu} \ ,
\eea
where $G$ is the field strength of some non-Abelian gauge group. Such a change can trigger a phase transition. Whether the phase transition is first order depends on other parameters such as number of colors and flavors. 
In this work, we point out that if the phase transition is of first-order, we may see an oscillatory pattern related to it.

\subsection{Dynamics of phase transition and bubble collision}

\subsubsection{Condition for first order phase transition to finish during inflation}

In this section, we give estimates of the energy scales and other parameters of the inflation and the spectator sector to generate first order phase transition while preserving the success of the inflation. We also estimate the size of the bubble, justifying the range assumed in the main text. 

Let's consider de Sitter inflation. The metric is 
\bea
d\tau^2 = dt^2 - e^{2Ht} (dx^2 + dy^2 + dz^2) \ .
\eea
The bubble nucleation rate per physical volume can be written as
\bea
\frac{\Gamma}{V_{\rm phy}} = C m_\sigma^4 e^{-S_4} \ ,
\eea
where $m_\sigma$ is the typical energy scale of the spectating sector, and $S_4$ is the bounce action. Therefore, the bubble nucleation rate per comoving volume at time $t$ can be written as
\bea
\frac{\Gamma}{V} = e^{3 H t} C m_\sigma^4 e^{-S_4} \ . 
\eea
Now, let's assume the bubbles expand with the speed of light. That means the point on the bubble wall evolves along a null geodesic curve. We have $dt = e^{Ht} dr$ for the bubble created at the origin of the space. Then, it is easy to see that for a bubble nucleated at $t'$, its comoving radius at $t$ can be written as
\bea
R(t,t') = \frac{1}{H} (e^{-H t'} - e^{-H t}) \ .
\eea
Then the fraction of the space that remains at the false vacuum at time $t$ can be written as~\cite{Guth:1981uk}
\bea
{\cal P} (t) &=& \exp\left[ - \int_{-\infty}^{t} dt' \frac{4 \pi}{3 H^3} \right.  \nn 
&& \times \left.(e^{-H t'} - e^{-H t})^3 e^{3H t'} C m_\sigma^4 e^{-S_4(t')} \right] \ .
\eea

Now, the necessary condition for the phase transition to finish at $t$ is that the exponential part of ${\cal P}(t)$ can achieve order of unity. Therefore we require 
\bea
\int_{-\infty}^{t}\!\! dt' \frac{4\pi}{3 H^3} (e^{-H t'} - e^{-H t})^3 e^{3H t'} C m_\sigma^4 e^{-S_4(t')}  \!\sim\! {\cal O}(1) \ . 
\eea
The bounce action $S_4$ at $t'$ can be expanded as 
\bea
S_4(t') = S_4(t) + \frac{d S_4(t)}{dt} (t' - t) \equiv S_4 (t) - \beta (t' - t) \ .
\eea 
Therefore, we have the condition for the phase transition to complete is 
\bea
{\cal O}(1) \!&\sim&\! C m_\sigma^4 e^{-S_4(t)} \frac{4\pi}{H^3} \int_{-\infty}^t d t' \left( 1 - e^{- H (t - t')}\right)^3 e^{-\beta(t - t')} \nn 
\!&\approx&\! C m_\sigma^4 e^{-S_4(t)} \frac{8\pi}{\beta (\beta+H)(\beta + 2H)(\beta + 3H) } \nn
\!&\approx&\! 8\pi C  e^{-S_4(t)} \frac{m_\sigma^4}{\beta^4} \ ,
\eea
where in the last step $\beta \gg H$ is assumed. Therefore, the requirement for first order phase transition to complete at $t_0$ is
\bea
S_4(t_0) \approx \log \left(\frac{m_\sigma^4}{ \beta^4}\right) \ .
\eea
The requirement that the phase transition is strong first order requires $S_4 \gg 1$, which indicates 
\bea\label{eq:12}
m_\sigma^4 \gg \beta^4 \ .
\eea
Whereas, on the other hand, the energy density in the spectating sector needs to be smaller than the total energy driven the inflation. We have 
\bea
m_\sigma^4 \ll M_{\rm pl}^2 H^2 \ .
\eea
Therefore, we need 
\bea
\left[\left(\frac{\beta}{H} \right)^4 H^2\right] H^2  \ll m_\sigma^4  \ll M_{\rm pl}^2 H^2 \ .
\eea

\begin{figure}[t!]
        \includegraphics[scale=0.6]{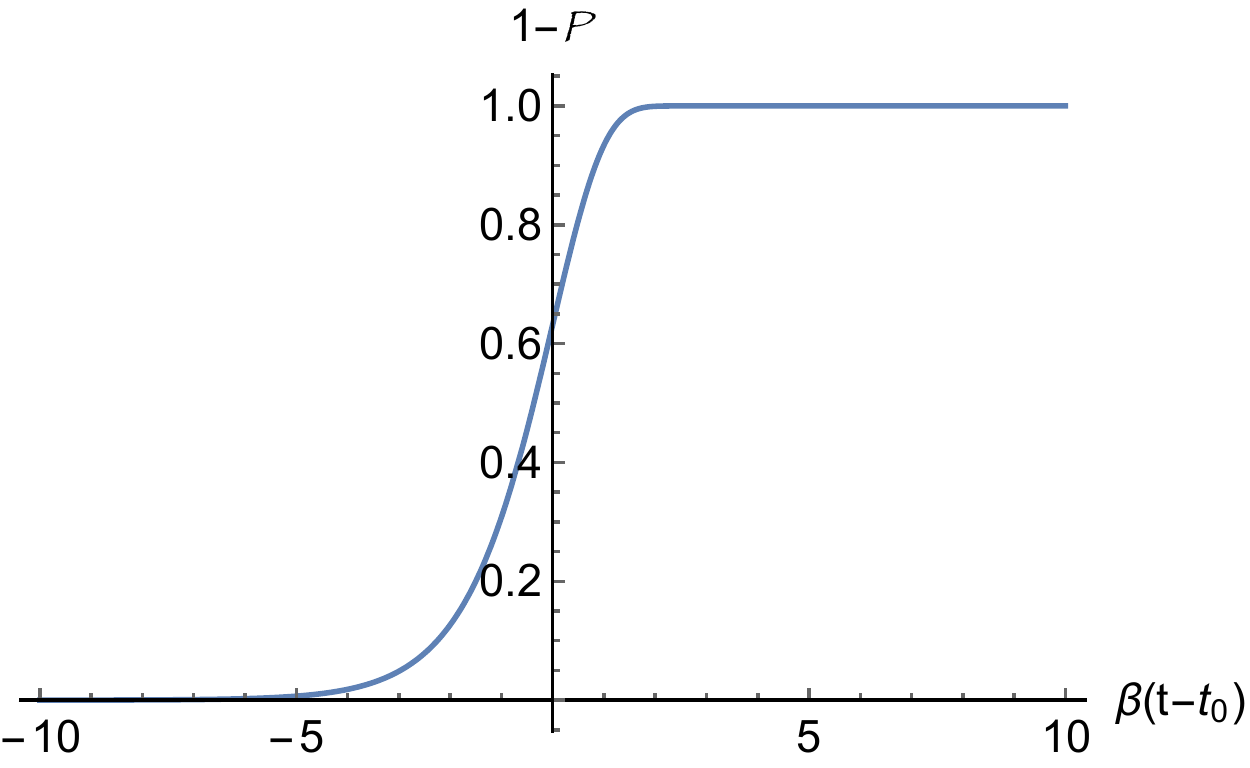}
\caption{Evolution of the fraction of the space occupied by the true vacuum. }
\label{fig:1}
\end{figure}

In the following we can see that the typical value of $\beta/H$ is about ${\cal O}(10) - {\cal O}(100)$. Therefore, for reasonable values of $H$ we always have 
\bea
\beta^4 \ll H^2 M_{\rm pl}^2 \ . 
\eea
As a result, we can always build model for first order phase transition to complete during inflation as long as the condition $\beta/ H \gg 1$ is fulfilled. The evolution of $1- {\cal P}$ is shown in Fig.~\ref{fig:1}, and one can see that as long as long as the condition $\beta/H \gg 1$ is fulfilled, the phase transition can complete within less then one e-fold.

\subsubsection{Typical values of $\beta/H$}

In a first order phase transition, the typical radius of the bubbles at the end of the phase transition can be estimated as $\beta^{-1} = |dS_4/dt|^{-1}$. We have
\bea
\beta &=& \left|\frac{d S_4}{d t}\right| = \left|\frac{d S_4}{d \mu_{\rm eff}^2}\right| \left|\frac{d\mu_{\rm eff}^2}{dt} \right|\nn
&=& \left|\frac{d S_4}{d\log \mu_{\rm eff}^2} \right| \left|\frac{d\mu_{\rm eff}^2}{\mu_{\rm eff}^2dt} \right|= I_1 S_4 \left|\frac{2 \dot\phi}{\phi - \frac{\mu^2}{c^2\phi}} \right| \ ,
\eea
where 
\bea
I_1 = \frac{1}{S_4} \left|\frac{d S_4}{d\log\mu_{\rm eff}^2} \right| \ ,
\eea
can be calculated numerically with 
CosmoTransition~\cite{Wainwright:2011kj}, and the results show that the value of $I_1$ can vary from 0.2 to 5. 


Therefore, we get
\bea
\frac{\beta}{H} = I_1 S_4 (2\epsilon)^{1/2} \times \frac{M_{\rm pl}}{|\phi - \frac{\mu^2}{c^2\phi}|} \ . 
\eea
In slow-roll single field models, we have the simple relation that~\cite{Baumann:2009ds}
\bea
\int_{\phi_{\rm end}}^{\phi_{\rm PT}} \frac{d\phi}{\sqrt{2\epsilon} M_{\rm pl}} = N_{\rm e} \ ,
\eea
where $N_{\rm e}$ is defined as the number of e-folds before the end of inflation. Therefore, if we use the range of $\phi$ to estimate the value of $\phi$ at the moment of the phase transition and assume epsilon does not evolve much after the phase transition, we have  
\bea\label{eq:11}
\frac{\beta}{H} \approx \frac{dS_4}{d\log\mu_{\rm eff}^2}\times \frac{1}{N_e|1 - \frac{\mu^2}{c^2 \phi^2}|} \ .
\eea
In the phase transition region, since it usually requires a small bounce, there is usually a cancelation between the two terms in $\mu_{\rm eff}^2$. For example, in the case of $V_1$ and $V_2$, during the phase transition, the ratio $\mu_{\rm eff}^2/\Lambda^2$ changes from order one to about $10\%$. Therefore, it is very probable that in the framework of slow-roll phase transition, the value of $\beta/H$ is about ${\cal O}$(10) to ${\cal O}$(100). 




\begin{figure}
\centering
\includegraphics[height=2.in]{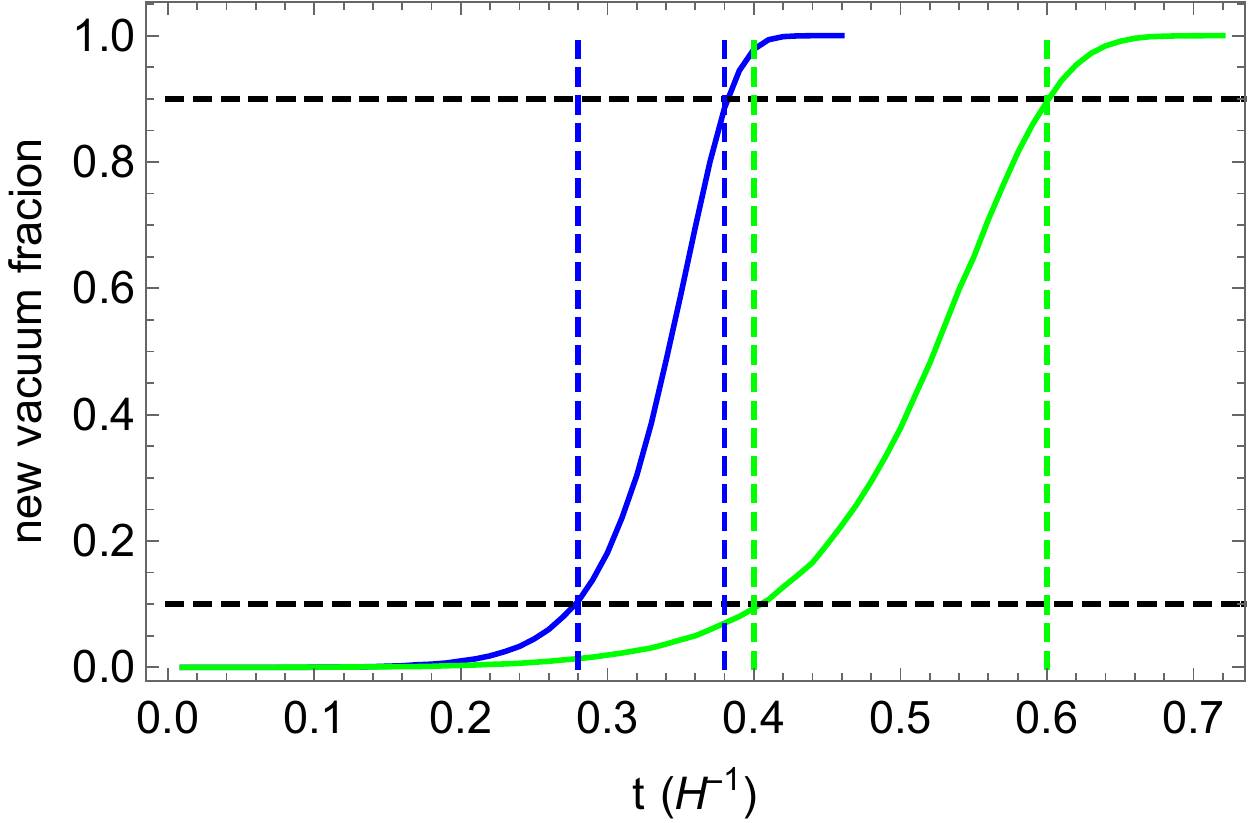}
\caption{Numerical simulation of the occupation fraction of the new vacuum for $\beta/H = 30$ (blue) and $\beta/H = 15$ (green). }\label{fig:numerical1}
\end{figure}

\begin{figure}
\centering
\includegraphics[height=2.in]{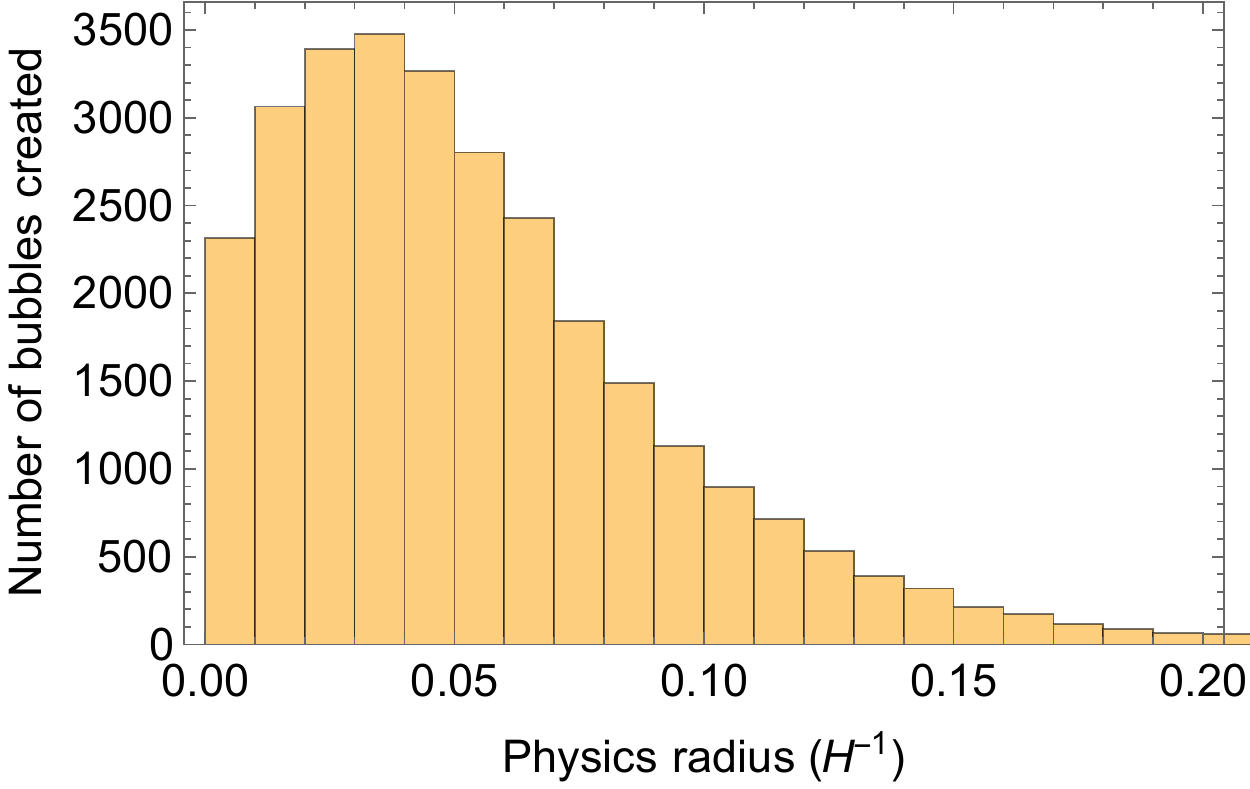}
\includegraphics[height=2.in]{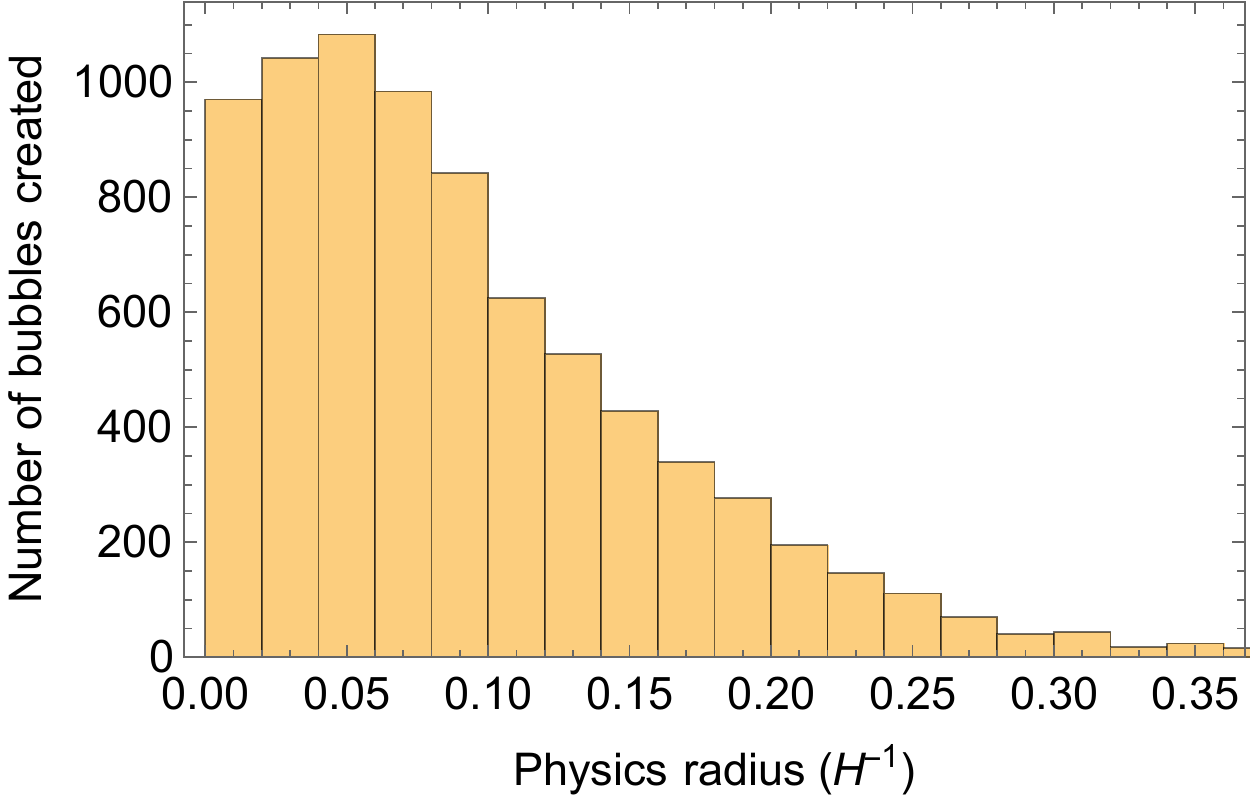}
\caption{Distributions of the physical radius of bubbles when the occupation faction of the new vacuum is about 90\% for $\beta/H = 30$ (left) and $\beta/H = 15$ (right). The vertical axis show the number of bubbles in a comoving volume of $8 H^{-3}$ in each radius bin. }\label{fig:radius}
\end{figure}

\begin{figure}
\centering
\includegraphics[height=2.5in]{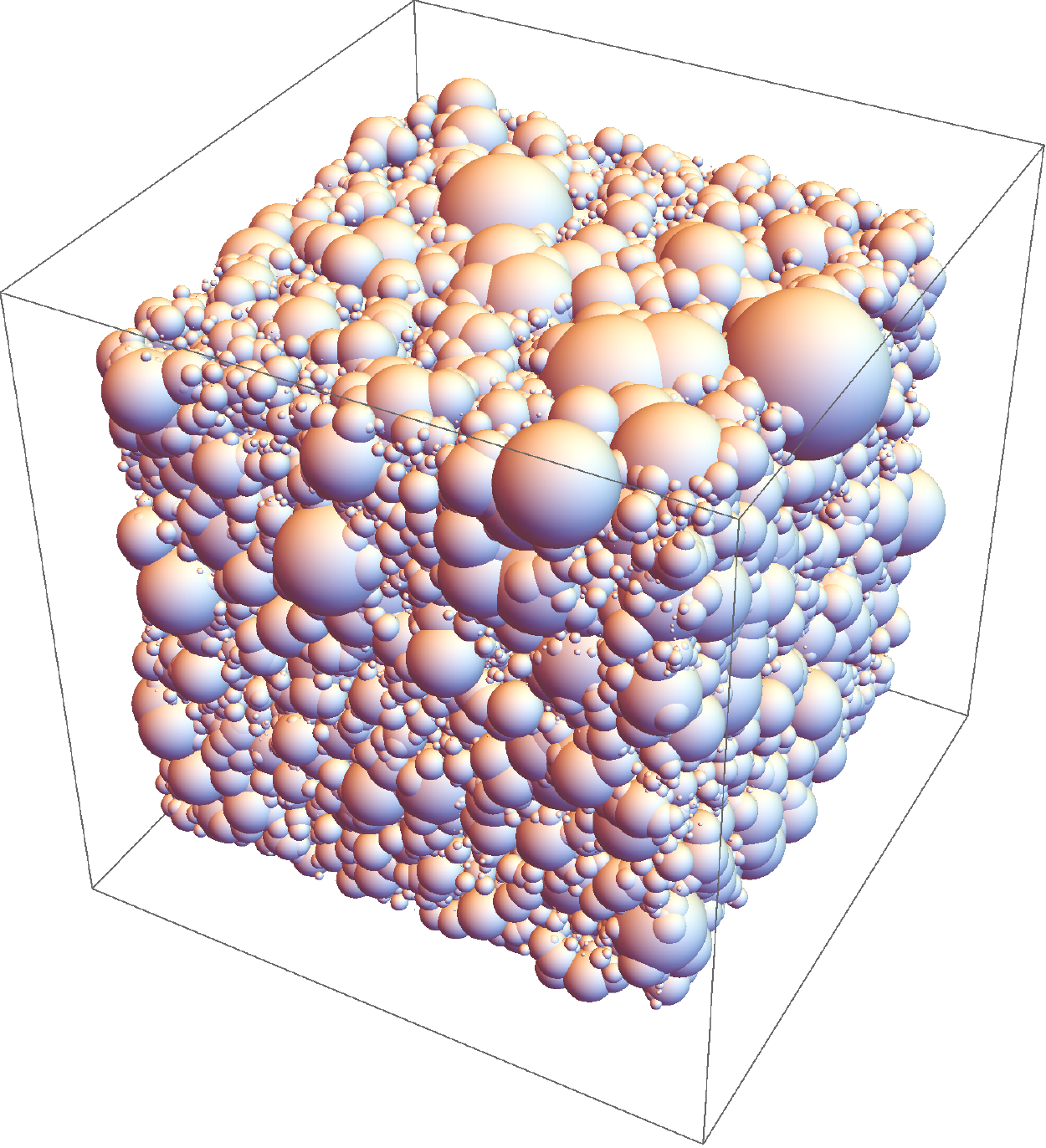}
\includegraphics[height=2.5in]{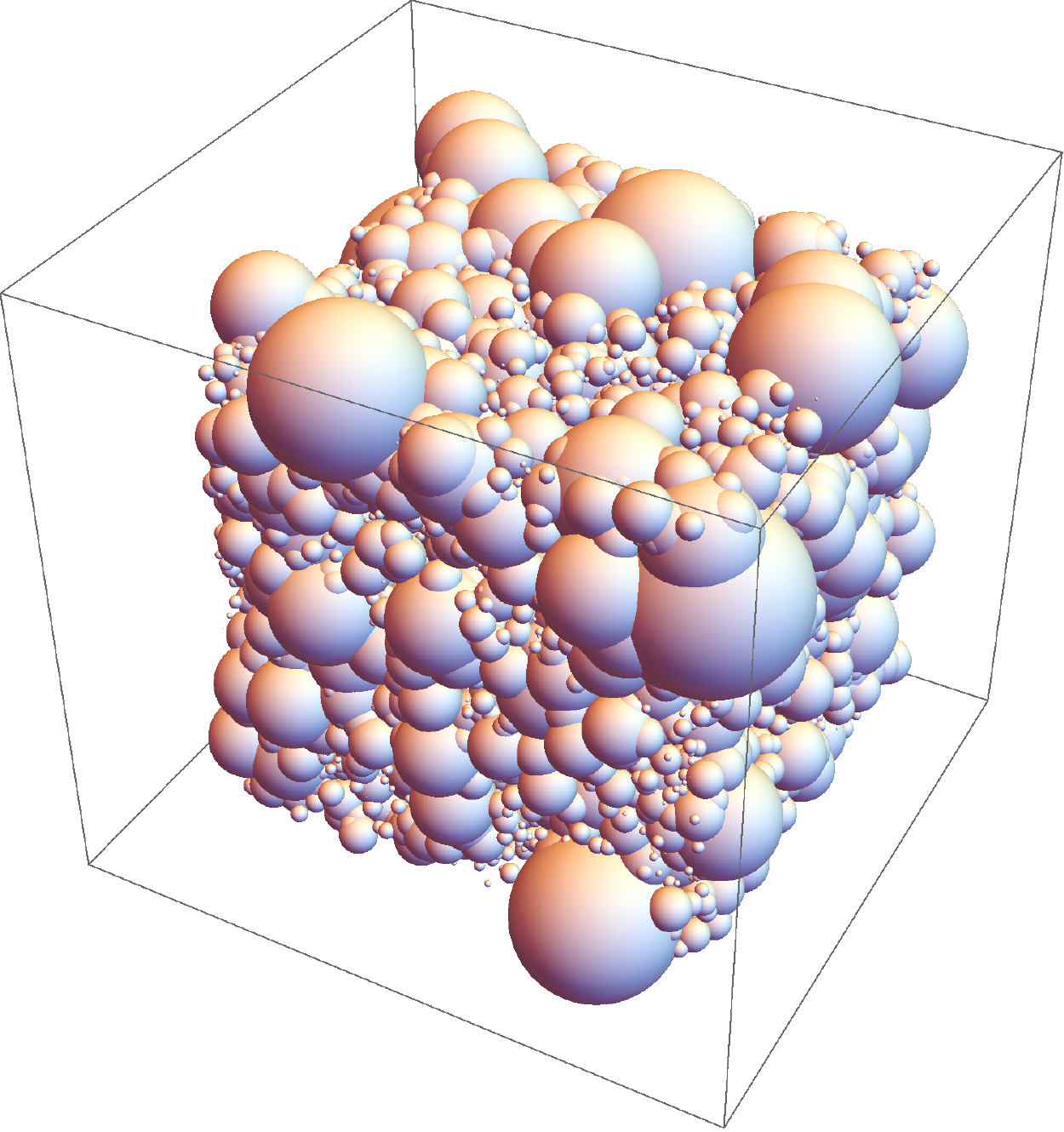}
\caption{Typical bubble configurations at the completion of the phase transition for $\beta/H = 30$ (left) and $\beta/H = 15$ (right) in the $2H^{-1}\times2H^{-1}\times 2H^{-1}$ comoving box.  }\label{fig:bubbles}
\end{figure}

\subsubsection{Numerical illustrations}

In order to illustrate the completeness of the phase transition, we did some numerical simulations of the expansion of the bubbles in de Sitter space. We start the simulation of bubble nucleation at $t_0$ when the we have $\Gamma/V_{\rm phy} = H^4$, namely there is one bubble nucleated at each Hubble patch in one e-fold. The condition that $S_4 \gg 1$ requires that $m_\sigma^4 \gg H^4$. This condition is weaker than Eq.~(\ref{eq:12}), and therefore can be fulfilled as discussed in the last subsection. Then, we expand $S_4$ as $S_4(t_0) - \beta(t - t_0)$. Then we randomly generate bubbles according to the nucleation rate. After nucleation, we assume that the bubbles expand with the speed of light. The evolution of the fraction occupied by the new vacuum is shown in Fig.~\ref{fig:numerical1}, where the blue and green curves are for $\beta/H = 30$ and 15. We can see that in both cases the time durations of the phase transition (from 10\% to 90\% as shown in Fig.~\ref{fig:numerical1}) are much smaller than $H^{-1}$. 

When the phase transition about to complete (e.g. when the occupation faction of the new vacuum is around 90\% at $t \approx 0.39 H^{-1}$ for $\beta/H = 30$ and $t\approx 0.6 H^{-1}$ for $\beta/H = 15$), the distribution of the physical radius of the bubbles are shown in Fig.~\ref{fig:radius}. One can see that the peak positions of the radius distributions in both cases are about $\beta^{-1}$. The bubble typical bubble configurations when the phase transition about to complete are shown in Fig.~\ref{fig:bubbles}. One can see that the size of the bubbles in the case of $\beta/H = 30$ is significantly smaller than the case of $\beta/H = 15$.

\subsection{Redshift of the signal strength}

Once the GW is generated, the subsequent evolution is a familiar story. In the main text, we quote the result of the solution of relevant equations of motion. Here, we give a scaling argument to understand the parametric dependence of the signal strength on the relevant scales of the model. 

For a GW mode with physical momentum $k_p$ at the moment it is produced. Its energy density redshifts with $a^{-4}$ before it exits the horizon when $k_p \sim H$. When it is outside the horizon before the end of inflation, the field value $h_{ij}$ as defined in the manuscript stays constant, while the momentum still redshifts as $a^{-1}$. As a result, in this period the GW energy density redshifts as $a^{-2}$. After the end of inflation, when the GW mode is still outside the horizon, for the same reason the energy density goes as $a^{-2}$. Once it evolves back inside the horizon, its energy density redshifts as $a^{-4}$ again. Therefore, the energy density of GW when it comes back into the horizon can be written as
\bea
\rho_{\rm GW}(t) \approx \rho_{\rm GW}^0 \times \left( \frac{H}{k_p} \right)^4 \left( \frac{a_1}{a_2} \right)^2 \left( \frac{a_2}{a_3} \right)^2 \left( \frac{a_3}{a(t)} \right)^4 \ ,
\eea
where $a_1$ is the scale factor when the mode evolves outside the horizon, $a_2$ when inflation ends, $a_3$ when the mode reenters the horizon. After inflation, in the radiation dominated era, the Hubble parameter is 
\bea 
H \sim \frac{T^2}{M_{\rm pl}} \sim a^{-2} \ .
\eea
Assuming $H$ is a constant during inflation. We have 
\bea
\frac{a_1}{a_2} = \frac{a_2}{a_3} \ .
\eea
The energy density can be rewritten as
\bea
\rho_{\rm GW}(t) &\approx& \rho_{\rm GW}^0 \times \left( \frac{H}{k_p} \right)^4  \left( \frac{a_2}{a_3} \right)^4 \left( \frac{a_3}{a(t)} \right)^4 \nn
&=& \rho_{\rm GW}^0 \times \left( \frac{H}{k_p} \right)^4  \left( \frac{a_2}{a(t)} \right)^4 \ .
\eea
In this work, we work in the case that most of the energy of the inflaton potential are converted into radiation. Hence, after inflation, the energy of the radiation evolves as 
\bea
\rho_{\rm R} \sim \rho_{\rm inf} \times \left(\frac{a_2}{a(t)}\right)^4 \ .
\eea
Finally,  today's abundance of GW is
\bea
\Omega_k = \Omega_R \times \frac{d\rho_{\rm GW}}{\rho_{R} d\log k } \sim \left( \frac{H}{k_p} \right)^4 \ . 
\eea
Therefore,  the GW signal is only diluted by a factor of $(H/k_p)^4$. This factor is explicitly shown in the ${\cal S}$ factor in Eq.~(10) of the manuscript. 
During phase transition, the typical value of $k_p$ is determined by the bubble radius $\beta$, given in (9) for a given model.

\bigskip

{\noindent \bf Acknowledgement}~~We thank Junwu Huang, Hongliang Jiang, Misao Sasaki, Wayne Hu, Yi Wang and Zhong-Zhi Xianyu for useful discussions. HA is supported by NSFC under Grant No. 11975134, the National Key Research and Development Program of China under Grant No.2017YFA0402204 and the Tsinghua University Initiative Scientific Research Program. KFL was supported in part by the National Science Foundation under Grant No. NSF PHY-1748958 and by the Heising-Simons Foundation and acknowledges the hospitality of Kavli Institute for Theoretical Physics while this work was in progress. LTW is supported by the DOE grant DE-SC0013642. The work of SZ was supported in part by the Swedish Research Council under grants number 2015-05333 and 2018-03803.

\bibliography{GW}

\begin{thebibliography}{52}
\expandafter\ifx\csname natexlab\endcsname\relax\def\natexlab#1{#1}\fi
\expandafter\ifx\csname bibnamefont\endcsname\relax
  \def\bibnamefont#1{#1}\fi
\expandafter\ifx\csname bibfnamefont\endcsname\relax
  \def\bibfnamefont#1{#1}\fi
\expandafter\ifx\csname citenamefont\endcsname\relax
  \def\citenamefont#1{#1}\fi
\expandafter\ifx\csname url\endcsname\relax
  \def\url#1{\texttt{#1}}\fi
\expandafter\ifx\csname urlprefix\endcsname\relax\def\urlprefix{URL }\fi
\providecommand{\bibinfo}[2]{#2}
\providecommand{\eprint}[2][]{\url{#2}}

\bibitem[{\citenamefont{Seoane et~al.}(2013)}]{Seoane:2013qna}
\bibinfo{author}{\bibfnamefont{P.~A.} \bibnamefont{Seoane}}
  \bibnamefont{et~al.} (\bibinfo{collaboration}{eLISA}) (\bibinfo{year}{2013}),
  \eprint{1305.5720}.

\bibitem[{\citenamefont{Amaro-Seoane et~al.}(2017)}]{Audley:2017drz}
\bibinfo{author}{\bibfnamefont{P.}~\bibnamefont{Amaro-Seoane}}
  \bibnamefont{et~al.} (\bibinfo{collaboration}{LISA}) (\bibinfo{year}{2017}),
  \eprint{1702.00786}.

\bibitem[{\citenamefont{Kawamura et~al.}(2011)}]{Kawamura:2011zz}
\bibinfo{author}{\bibfnamefont{S.}~\bibnamefont{Kawamura}}
  \bibnamefont{et~al.}, \bibinfo{journal}{Class. Quant. Grav.}
  \textbf{\bibinfo{volume}{28}}, \bibinfo{pages}{094011}
  (\bibinfo{year}{2011}).

\bibitem[{\citenamefont{Luo et~al.}(2016)}]{Luo:2015ght}
\bibinfo{author}{\bibfnamefont{J.}~\bibnamefont{Luo}} \bibnamefont{et~al.}
  (\bibinfo{collaboration}{TianQin}), \bibinfo{journal}{Class. Quant. Grav.}
  \textbf{\bibinfo{volume}{33}}, \bibinfo{pages}{035010}
  (\bibinfo{year}{2016}), \eprint{1512.02076}.

\bibitem[{\citenamefont{Ruan et~al.}(2020)\citenamefont{Ruan, Guo, Cai, and
  Zhang}}]{Guo:2018npi}
\bibinfo{author}{\bibfnamefont{W.-H.} \bibnamefont{Ruan}},
  \bibinfo{author}{\bibfnamefont{Z.-K.} \bibnamefont{Guo}},
  \bibinfo{author}{\bibfnamefont{R.-G.} \bibnamefont{Cai}}, \bibnamefont{and}
  \bibinfo{author}{\bibfnamefont{Y.-Z.} \bibnamefont{Zhang}},
  \bibinfo{journal}{Int. J. Mod. Phys. A} \textbf{\bibinfo{volume}{35}},
  \bibinfo{pages}{2050075} (\bibinfo{year}{2020}), \eprint{1807.09495}.

\bibitem[{\citenamefont{Crowder and Cornish}(2005)}]{Crowder:2005nr}
\bibinfo{author}{\bibfnamefont{J.}~\bibnamefont{Crowder}} \bibnamefont{and}
  \bibinfo{author}{\bibfnamefont{N.~J.} \bibnamefont{Cornish}},
  \bibinfo{journal}{Phys. Rev. D} \textbf{\bibinfo{volume}{72}},
  \bibinfo{pages}{083005} (\bibinfo{year}{2005}), \eprint{gr-qc/0506015}.

\bibitem[{\citenamefont{Harry et~al.}(2006)\citenamefont{Harry, Fritschel,
  Shaddock, Folkner, and Phinney}}]{Harry:2006fi}
\bibinfo{author}{\bibfnamefont{G.}~\bibnamefont{Harry}},
  \bibinfo{author}{\bibfnamefont{P.}~\bibnamefont{Fritschel}},
  \bibinfo{author}{\bibfnamefont{D.}~\bibnamefont{Shaddock}},
  \bibinfo{author}{\bibfnamefont{W.}~\bibnamefont{Folkner}}, \bibnamefont{and}
  \bibinfo{author}{\bibfnamefont{E.}~\bibnamefont{Phinney}},
  \bibinfo{journal}{Class. Quant. Grav.} \textbf{\bibinfo{volume}{23}},
  \bibinfo{pages}{4887} (\bibinfo{year}{2006}), \bibinfo{note}{[Erratum:
  Class.Quant.Grav. 23, 7361 (2006)]}.

\bibitem[{\citenamefont{Corbin and Cornish}(2006)}]{Corbin:2005ny}
\bibinfo{author}{\bibfnamefont{V.}~\bibnamefont{Corbin}} \bibnamefont{and}
  \bibinfo{author}{\bibfnamefont{N.~J.} \bibnamefont{Cornish}},
  \bibinfo{journal}{Class. Quant. Grav.} \textbf{\bibinfo{volume}{23}},
  \bibinfo{pages}{2435} (\bibinfo{year}{2006}), \eprint{gr-qc/0512039}.

\bibitem[{\citenamefont{Kramer and Champion}(2013)}]{Kramer:2013kea}
\bibinfo{author}{\bibfnamefont{M.}~\bibnamefont{Kramer}} \bibnamefont{and}
  \bibinfo{author}{\bibfnamefont{D.~J.} \bibnamefont{Champion}},
  \bibinfo{journal}{Class. Quant. Grav.} \textbf{\bibinfo{volume}{30}},
  \bibinfo{pages}{224009} (\bibinfo{year}{2013}).

\bibitem[{\citenamefont{Hobbs et~al.}(2010)}]{Hobbs:2009yy}
\bibinfo{author}{\bibfnamefont{G.}~\bibnamefont{Hobbs}} \bibnamefont{et~al.},
  \bibinfo{journal}{Class. Quant. Grav.} \textbf{\bibinfo{volume}{27}},
  \bibinfo{pages}{084013} (\bibinfo{year}{2010}), \eprint{0911.5206}.

\bibitem[{\citenamefont{Janssen et~al.}(2015)}]{Janssen:2014dka}
\bibinfo{author}{\bibfnamefont{G.}~\bibnamefont{Janssen}} \bibnamefont{et~al.},
  \bibinfo{journal}{PoS} \textbf{\bibinfo{volume}{AASKA14}},
  \bibinfo{pages}{037} (\bibinfo{year}{2015}), \eprint{1501.00127}.

\bibitem[{\citenamefont{Aasi et~al.}(2015)}]{TheLIGOScientific:2014jea}
\bibinfo{author}{\bibfnamefont{J.}~\bibnamefont{Aasi}} \bibnamefont{et~al.}
  (\bibinfo{collaboration}{LIGO Scientific}), \bibinfo{journal}{Class. Quant.
  Grav.} \textbf{\bibinfo{volume}{32}}, \bibinfo{pages}{074001}
  (\bibinfo{year}{2015}), \eprint{1411.4547}.

\bibitem[{\citenamefont{Abramovici et~al.}(1992)}]{Abramovici:1992ah}
\bibinfo{author}{\bibfnamefont{A.}~\bibnamefont{Abramovici}}
  \bibnamefont{et~al.}, \bibinfo{journal}{Science}
  \textbf{\bibinfo{volume}{256}}, \bibinfo{pages}{325} (\bibinfo{year}{1992}).

\bibitem[{\citenamefont{Acernese et~al.}(2015)}]{TheVirgo:2014hva}
\bibinfo{author}{\bibfnamefont{F.}~\bibnamefont{Acernese}} \bibnamefont{et~al.}
  (\bibinfo{collaboration}{VIRGO}), \bibinfo{journal}{Class. Quant. Grav.}
  \textbf{\bibinfo{volume}{32}}, \bibinfo{pages}{024001}
  (\bibinfo{year}{2015}), \eprint{1408.3978}.

\bibitem[{\citenamefont{Punturo et~al.}(2010)}]{Punturo:2010zz}
\bibinfo{author}{\bibfnamefont{M.}~\bibnamefont{Punturo}} \bibnamefont{et~al.},
  \bibinfo{journal}{Class. Quant. Grav.} \textbf{\bibinfo{volume}{27}},
  \bibinfo{pages}{194002} (\bibinfo{year}{2010}).

\bibitem[{\citenamefont{Reitze et~al.}(2019)}]{Reitze:2019iox}
\bibinfo{author}{\bibfnamefont{D.}~\bibnamefont{Reitze}} \bibnamefont{et~al.},
  \bibinfo{journal}{Bull. Am. Astron. Soc.} \textbf{\bibinfo{volume}{51}},
  \bibinfo{pages}{035} (\bibinfo{year}{2019}), \eprint{1907.04833}.

\bibitem[{\citenamefont{Hui et~al.}(2018)}]{Hui:2018cvg}
\bibinfo{author}{\bibfnamefont{H.}~\bibnamefont{Hui}} \bibnamefont{et~al.},
  \bibinfo{journal}{Proc. SPIE Int. Soc. Opt. Eng.}
  \textbf{\bibinfo{volume}{10708}}, \bibinfo{pages}{1070807}
  (\bibinfo{year}{2018}), \eprint{1808.00568}.

\bibitem[{\citenamefont{Li et~al.}(2019)}]{Li:2017drr}
\bibinfo{author}{\bibfnamefont{H.}~\bibnamefont{Li}} \bibnamefont{et~al.},
  \bibinfo{journal}{Natl. Sci. Rev.} \textbf{\bibinfo{volume}{6}},
  \bibinfo{pages}{145} (\bibinfo{year}{2019}), \eprint{1710.03047}.

\bibitem[{\citenamefont{Abazajian et~al.}(2019)}]{Abazajian:2019eic}
\bibinfo{author}{\bibfnamefont{K.}~\bibnamefont{Abazajian}}
  \bibnamefont{et~al.} (\bibinfo{year}{2019}), \eprint{1907.04473}.

\bibitem[{\citenamefont{Grishchuk}(1975)}]{Grishchuk:1974ny}
\bibinfo{author}{\bibfnamefont{L.}~\bibnamefont{Grishchuk}},
  \bibinfo{journal}{Sov. Phys. JETP} \textbf{\bibinfo{volume}{40}},
  \bibinfo{pages}{409} (\bibinfo{year}{1975}).

\bibitem[{\citenamefont{Starobinsky}(1979)}]{Starobinsky:1979ty}
\bibinfo{author}{\bibfnamefont{A.~A.} \bibnamefont{Starobinsky}},
  \bibinfo{journal}{JETP Lett.} \textbf{\bibinfo{volume}{30}},
  \bibinfo{pages}{682} (\bibinfo{year}{1979}).

\bibitem[{\citenamefont{Rubakov et~al.}(1982)\citenamefont{Rubakov, Sazhin, and
  Veryaskin}}]{Rubakov:1982df}
\bibinfo{author}{\bibfnamefont{V.}~\bibnamefont{Rubakov}},
  \bibinfo{author}{\bibfnamefont{M.}~\bibnamefont{Sazhin}}, \bibnamefont{and}
  \bibinfo{author}{\bibfnamefont{A.}~\bibnamefont{Veryaskin}},
  \bibinfo{journal}{Phys. Lett. B} \textbf{\bibinfo{volume}{115}},
  \bibinfo{pages}{189} (\bibinfo{year}{1982}).

\bibitem[{\citenamefont{Fabbri and Pollock}(1983)}]{Fabbri:1983us}
\bibinfo{author}{\bibfnamefont{R.}~\bibnamefont{Fabbri}} \bibnamefont{and}
  \bibinfo{author}{\bibfnamefont{M.}~\bibnamefont{Pollock}},
  \bibinfo{journal}{Phys. Lett. B} \textbf{\bibinfo{volume}{125}},
  \bibinfo{pages}{445} (\bibinfo{year}{1983}).

\bibitem[{\citenamefont{Abbott and Wise}(1984)}]{Abbott:1984fp}
\bibinfo{author}{\bibfnamefont{L.}~\bibnamefont{Abbott}} \bibnamefont{and}
  \bibinfo{author}{\bibfnamefont{M.~B.} \bibnamefont{Wise}},
  \bibinfo{journal}{Nucl. Phys. B} \textbf{\bibinfo{volume}{244}},
  \bibinfo{pages}{541} (\bibinfo{year}{1984}).

\bibitem[{\citenamefont{Witten}(1984)}]{Witten:1984rs}
\bibinfo{author}{\bibfnamefont{E.}~\bibnamefont{Witten}},
  \bibinfo{journal}{Phys. Rev. D} \textbf{\bibinfo{volume}{30}},
  \bibinfo{pages}{272} (\bibinfo{year}{1984}).

\bibitem[{\citenamefont{Kamionkowski et~al.}(1994)\citenamefont{Kamionkowski,
  Kosowsky, and Turner}}]{Kamionkowski:1993fg}
\bibinfo{author}{\bibfnamefont{M.}~\bibnamefont{Kamionkowski}},
  \bibinfo{author}{\bibfnamefont{A.}~\bibnamefont{Kosowsky}}, \bibnamefont{and}
  \bibinfo{author}{\bibfnamefont{M.~S.} \bibnamefont{Turner}},
  \bibinfo{journal}{Phys. Rev. D} \textbf{\bibinfo{volume}{49}},
  \bibinfo{pages}{2837} (\bibinfo{year}{1994}), \eprint{astro-ph/9310044}.

\bibitem[{\citenamefont{Vachaspati and Vilenkin}(1985)}]{Vachaspati:1984gt}
\bibinfo{author}{\bibfnamefont{T.}~\bibnamefont{Vachaspati}} \bibnamefont{and}
  \bibinfo{author}{\bibfnamefont{A.}~\bibnamefont{Vilenkin}},
  \bibinfo{journal}{Phys. Rev. D} \textbf{\bibinfo{volume}{31}},
  \bibinfo{pages}{3052} (\bibinfo{year}{1985}).

\bibitem[{\citenamefont{Brandenberger et~al.}(1986)\citenamefont{Brandenberger,
  Albrecht, and Turok}}]{Brandenberger:1986xn}
\bibinfo{author}{\bibfnamefont{R.~H.} \bibnamefont{Brandenberger}},
  \bibinfo{author}{\bibfnamefont{A.}~\bibnamefont{Albrecht}}, \bibnamefont{and}
  \bibinfo{author}{\bibfnamefont{N.}~\bibnamefont{Turok}},
  \bibinfo{journal}{Nucl. Phys. B} \textbf{\bibinfo{volume}{277}},
  \bibinfo{pages}{605} (\bibinfo{year}{1986}).

\bibitem[{\citenamefont{Hindmarsh}(1990)}]{Hindmarsh:1990xi}
\bibinfo{author}{\bibfnamefont{M.}~\bibnamefont{Hindmarsh}},
  \bibinfo{journal}{Phys. Lett. B} \textbf{\bibinfo{volume}{251}},
  \bibinfo{pages}{28} (\bibinfo{year}{1990}).

\bibitem[{\citenamefont{Damour and Vilenkin}(2001)}]{Damour:2001bk}
\bibinfo{author}{\bibfnamefont{T.}~\bibnamefont{Damour}} \bibnamefont{and}
  \bibinfo{author}{\bibfnamefont{A.}~\bibnamefont{Vilenkin}},
  \bibinfo{journal}{Phys. Rev. D} \textbf{\bibinfo{volume}{64}},
  \bibinfo{pages}{064008} (\bibinfo{year}{2001}), \eprint{gr-qc/0104026}.

\bibitem[{\citenamefont{Siemens and Olum}(2001)}]{Siemens:2001dx}
\bibinfo{author}{\bibfnamefont{X.}~\bibnamefont{Siemens}} \bibnamefont{and}
  \bibinfo{author}{\bibfnamefont{K.~D.} \bibnamefont{Olum}},
  \bibinfo{journal}{Nucl. Phys. B} \textbf{\bibinfo{volume}{611}},
  \bibinfo{pages}{125} (\bibinfo{year}{2001}), \bibinfo{note}{[Erratum:
  Nucl.Phys.B 645, 367--367 (2002)]}, \eprint{gr-qc/0104085}.

\bibitem[{\citenamefont{Hindmarsh and Kibble}(1995)}]{Hindmarsh:1994re}
\bibinfo{author}{\bibfnamefont{M.}~\bibnamefont{Hindmarsh}} \bibnamefont{and}
  \bibinfo{author}{\bibfnamefont{T.}~\bibnamefont{Kibble}},
  \bibinfo{journal}{Rept. Prog. Phys.} \textbf{\bibinfo{volume}{58}},
  \bibinfo{pages}{477} (\bibinfo{year}{1995}), \eprint{hep-ph/9411342}.

\bibitem[{\citenamefont{Guth}(1987)}]{Guth:1980zm}
\bibinfo{author}{\bibfnamefont{A.~H.} \bibnamefont{Guth}},
  \bibinfo{journal}{Adv. Ser. Astrophys. Cosmol.} \textbf{\bibinfo{volume}{3}},
  \bibinfo{pages}{139} (\bibinfo{year}{1987}).

\bibitem[{\citenamefont{Linde}(1987)}]{Linde:1981mu}
\bibinfo{author}{\bibfnamefont{A.~D.} \bibnamefont{Linde}},
  \bibinfo{journal}{Adv. Ser. Astrophys. Cosmol.} \textbf{\bibinfo{volume}{3}},
  \bibinfo{pages}{149} (\bibinfo{year}{1987}).

\bibitem[{\citenamefont{Albrecht and Steinhardt}(1987)}]{Albrecht:1982wi}
\bibinfo{author}{\bibfnamefont{A.}~\bibnamefont{Albrecht}} \bibnamefont{and}
  \bibinfo{author}{\bibfnamefont{P.~J.} \bibnamefont{Steinhardt}},
  \bibinfo{journal}{Adv. Ser. Astrophys. Cosmol.} \textbf{\bibinfo{volume}{3}},
  \bibinfo{pages}{158} (\bibinfo{year}{1987}).

\bibitem[{\citenamefont{Baumann}(2011)}]{Baumann:2009ds}
\bibinfo{author}{\bibfnamefont{D.}~\bibnamefont{Baumann}}, in
  \emph{\bibinfo{booktitle}{{Theoretical Advanced Study Institute in Elementary
  Particle Physics}: {Physics of the Large and the Small}}}
  (\bibinfo{year}{2011}), pp. \bibinfo{pages}{523--686}, \eprint{0907.5424}.

\bibitem[{\citenamefont{Chen and Wang}(2010)}]{Chen:2009zp}
\bibinfo{author}{\bibfnamefont{X.}~\bibnamefont{Chen}} \bibnamefont{and}
  \bibinfo{author}{\bibfnamefont{Y.}~\bibnamefont{Wang}},
  \bibinfo{journal}{JCAP} \textbf{\bibinfo{volume}{04}}, \bibinfo{pages}{027}
  (\bibinfo{year}{2010}), \eprint{0911.3380}.

\bibitem[{\citenamefont{Berera and Fang}(1995)}]{Berera:1995wh}
\bibinfo{author}{\bibfnamefont{A.}~\bibnamefont{Berera}} \bibnamefont{and}
  \bibinfo{author}{\bibfnamefont{L.-Z.} \bibnamefont{Fang}},
  \bibinfo{journal}{Phys. Rev. Lett.} \textbf{\bibinfo{volume}{74}},
  \bibinfo{pages}{1912} (\bibinfo{year}{1995}), \eprint{astro-ph/9501024}.

\bibitem[{\citenamefont{Berera}(1995)}]{Berera:1995ie}
\bibinfo{author}{\bibfnamefont{A.}~\bibnamefont{Berera}},
  \bibinfo{journal}{Phys. Rev. Lett.} \textbf{\bibinfo{volume}{75}},
  \bibinfo{pages}{3218} (\bibinfo{year}{1995}), \eprint{astro-ph/9509049}.

\bibitem[{\citenamefont{Jiang et~al.}(2017)\citenamefont{Jiang, Liu, Sun, and
  Wang}}]{Jiang:2015qor}
\bibinfo{author}{\bibfnamefont{H.}~\bibnamefont{Jiang}},
  \bibinfo{author}{\bibfnamefont{T.}~\bibnamefont{Liu}},
  \bibinfo{author}{\bibfnamefont{S.}~\bibnamefont{Sun}}, \bibnamefont{and}
  \bibinfo{author}{\bibfnamefont{Y.}~\bibnamefont{Wang}},
  \bibinfo{journal}{Phys. Lett. B} \textbf{\bibinfo{volume}{765}},
  \bibinfo{pages}{339} (\bibinfo{year}{2017}), \eprint{1512.07538}.

\bibitem[{\citenamefont{Wang et~al.}(2019)\citenamefont{Wang, Cai, and
  Piao}}]{Wang:2018caj}
\bibinfo{author}{\bibfnamefont{Y.-T.} \bibnamefont{Wang}},
  \bibinfo{author}{\bibfnamefont{Y.}~\bibnamefont{Cai}}, \bibnamefont{and}
  \bibinfo{author}{\bibfnamefont{Y.-S.} \bibnamefont{Piao}},
  \bibinfo{journal}{Phys. Lett. B} \textbf{\bibinfo{volume}{789}},
  \bibinfo{pages}{191} (\bibinfo{year}{2019}), \eprint{1801.03639}.

\bibitem[{\citenamefont{Sugimura et~al.}(2012)\citenamefont{Sugimura, Yamauchi,
  and Sasaki}}]{Sugimura:2011tk}
\bibinfo{author}{\bibfnamefont{K.}~\bibnamefont{Sugimura}},
  \bibinfo{author}{\bibfnamefont{D.}~\bibnamefont{Yamauchi}}, \bibnamefont{and}
  \bibinfo{author}{\bibfnamefont{M.}~\bibnamefont{Sasaki}},
  \bibinfo{journal}{JCAP} \textbf{\bibinfo{volume}{01}}, \bibinfo{pages}{027}
  (\bibinfo{year}{2012}), \eprint{1110.4773}.

\bibitem[{\citenamefont{Cai et~al.}(2019)\citenamefont{Cai, Pi, and
  Sasaki}}]{Cai:2019cdl}
\bibinfo{author}{\bibfnamefont{R.-G.} \bibnamefont{Cai}},
  \bibinfo{author}{\bibfnamefont{S.}~\bibnamefont{Pi}}, \bibnamefont{and}
  \bibinfo{author}{\bibfnamefont{M.}~\bibnamefont{Sasaki}}
  (\bibinfo{year}{2019}), \eprint{1909.13728}.

\bibitem[{\citenamefont{Caprini et~al.}(2009)\citenamefont{Caprini, Durrer,
  Konstandin, and Servant}}]{Caprini:2009fx}
\bibinfo{author}{\bibfnamefont{C.}~\bibnamefont{Caprini}},
  \bibinfo{author}{\bibfnamefont{R.}~\bibnamefont{Durrer}},
  \bibinfo{author}{\bibfnamefont{T.}~\bibnamefont{Konstandin}},
  \bibnamefont{and} \bibinfo{author}{\bibfnamefont{G.}~\bibnamefont{Servant}},
  \bibinfo{journal}{Phys. Rev. D} \textbf{\bibinfo{volume}{79}},
  \bibinfo{pages}{083519} (\bibinfo{year}{2009}), \eprint{0901.1661}.

\bibitem[{\citenamefont{Huber and Konstandin}(2008)}]{Huber:2008hg}
\bibinfo{author}{\bibfnamefont{S.~J.} \bibnamefont{Huber}} \bibnamefont{and}
  \bibinfo{author}{\bibfnamefont{T.}~\bibnamefont{Konstandin}},
  \bibinfo{journal}{JCAP} \textbf{\bibinfo{volume}{09}}, \bibinfo{pages}{022}
  (\bibinfo{year}{2008}), \eprint{0806.1828}.

\bibitem[{\citenamefont{Harry}(2009)}]{Harry:BBO2}
\bibinfo{author}{\bibfnamefont{G.}~\bibnamefont{Harry}},
  \bibinfo{journal}{https://dcc.ligo.org/public/0002/G0900426/001
  /G0900426-v1.pdf}  (\bibinfo{year}{2009}).

\bibitem[{\citenamefont{Moore et~al.}(2015)\citenamefont{Moore, Cole, and
  Berry}}]{Moore:2014lga}
\bibinfo{author}{\bibfnamefont{C.}~\bibnamefont{Moore}},
  \bibinfo{author}{\bibfnamefont{R.}~\bibnamefont{Cole}}, \bibnamefont{and}
  \bibinfo{author}{\bibfnamefont{C.}~\bibnamefont{Berry}},
  \bibinfo{journal}{Class. Quant. Grav.} \textbf{\bibinfo{volume}{32}},
  \bibinfo{pages}{015014} (\bibinfo{year}{2015}), \eprint{1408.0740}.

\bibitem[{\citenamefont{Ade et~al.}(2016)}]{Array:2015xqh}
\bibinfo{author}{\bibfnamefont{P.}~\bibnamefont{Ade}} \bibnamefont{et~al.}
  (\bibinfo{collaboration}{BICEP2, Keck Array}), \bibinfo{journal}{Phys. Rev.
  Lett.} \textbf{\bibinfo{volume}{116}}, \bibinfo{pages}{031302}
  (\bibinfo{year}{2016}), \eprint{1510.09217}.

\bibitem[{\citenamefont{Blas et~al.}(2011)\citenamefont{Blas, Lesgourgues, and
  Tram}}]{Blas:2011rf}
\bibinfo{author}{\bibfnamefont{D.}~\bibnamefont{Blas}},
  \bibinfo{author}{\bibfnamefont{J.}~\bibnamefont{Lesgourgues}},
  \bibnamefont{and} \bibinfo{author}{\bibfnamefont{T.}~\bibnamefont{Tram}},
  \bibinfo{journal}{JCAP} \textbf{\bibinfo{volume}{07}}, \bibinfo{pages}{034}
  (\bibinfo{year}{2011}), \eprint{1104.2933}.

\bibitem[{\citenamefont{Chung et~al.}(2013)\citenamefont{Chung, Long, and
  Wang}}]{Chung:2012vg}
\bibinfo{author}{\bibfnamefont{D.~J.~H.} \bibnamefont{Chung}},
  \bibinfo{author}{\bibfnamefont{A.~J.} \bibnamefont{Long}}, \bibnamefont{and}
  \bibinfo{author}{\bibfnamefont{L.-T.} \bibnamefont{Wang}},
  \bibinfo{journal}{Phys. Rev. D} \textbf{\bibinfo{volume}{87}},
  \bibinfo{pages}{023509} (\bibinfo{year}{2013}), \eprint{1209.1819}.

\bibitem[{\citenamefont{Guth and Weinberg}(1981)}]{Guth:1981uk}
\bibinfo{author}{\bibfnamefont{A.~H.} \bibnamefont{Guth}} \bibnamefont{and}
  \bibinfo{author}{\bibfnamefont{E.~J.} \bibnamefont{Weinberg}},
  \bibinfo{journal}{Phys. Rev. D} \textbf{\bibinfo{volume}{23}},
  \bibinfo{pages}{876} (\bibinfo{year}{1981}).

\bibitem[{\citenamefont{Wainwright}(2012)}]{Wainwright:2011kj}
\bibinfo{author}{\bibfnamefont{C.~L.} \bibnamefont{Wainwright}},
  \bibinfo{journal}{Comput. Phys. Commun.} \textbf{\bibinfo{volume}{183}},
  \bibinfo{pages}{2006} (\bibinfo{year}{2012}), \eprint{1109.4189}.

\end{thebibliography}

\end{document}